\documentclass[aps,prc,amsmath,preprint,superscriptaddress,showkeys,showpacs]{revtex4-1}
\usepackage{graphicx}
\usepackage{epsfig,epstopdf}
\newcommand{\beqy}{\begin{eqnarray}}
\newcommand{\eeqy}{\end{eqnarray}}

\begin{document}

\title{Electron exchange and polarization effects on electron captures and neutron emissions by nuclei in white dwarfs and neutron stars}
\author{N. Chamel}
\affiliation{Institut d'Astronomie et d'Astrophysique, CP-226, Universit\'e Libre de Bruxelles,
1050 Brussels, Belgium}
\author{A.~F. Fantina}
\affiliation{Institut d'Astronomie et d'Astrophysique, CP-226, Universit\'e Libre de Bruxelles,
1050 Brussels, Belgium}
\affiliation{Grand Acc\'el\'erateur National d'Ions Lourds (GANIL), CEA/DSM - CNRS/IN2P3, Bvd Henri Becquerel, BP 55027, F-14076 Caen Cedex 5, France}

\date{\today}

\begin{abstract}
In dense stellar environments, nuclei may become unstable against electron captures and/or neutron emissions. These processes are 
of particular importance for determining the internal constitution of white-dwarf cores and neutron-star crusts. In this paper, the 
role of electron exchange and polarization effects is studied. In particular, the instability condition for the onset of electron captures 
and neutron emissions is extended so as to account for electron exchange and polarization. Moreover, general analytical expressions for the 
corresponding density and pressure are derived. The corrections to the electron-capture threshold in white-dwarf cores are found to be very 
small. Likewise, the neutron-drip density and pressure in the crusts of accreting and nonaccreting neutron stars are only slightly shifted. 
Depending on the nuclear mass model employed, electron polarization may change the composition of the crust of nonaccreting neutron stars. On
the other hand, the current uncertainties in the masses of neutron-rich Kr and Sr isotopes are found to be more important than electron exchange 
and polarization effects.
\end{abstract}



\maketitle

\section{introduction}
\label{intro} 

Electron captures and neutron emissions by atomic nuclei are among the most important processes governing the late evolution of stars
(see, e.g., Ref.~\cite{lang2014} for a recent review). In the dense core of white dwarfs, the onset of electron captures leads to 
a softening of the equation of state: as electrons combine with nuclei, further compression of matter does not increase the pressure. 
For this reason, electron captures limit the maximum possible mass of white dwarfs (depending on the core composition, the maximum mass 
may be further limited by general relativity, see e.g. Ref.\cite{shapiro1983}). Electron captures are also responsible for the production   
of very neutron-rich nuclei in the outer crust of a neutron star (see, e.g., Refs.~\cite{bps1971,roca2008,pearson2011,wolf2013,kreim2013}). Deeper 
in the crust, electron captures accompanied by neutron emissions lead to the appearance of a neutron liquid (see, e.g., Ref.~\cite{lrr}).

We have recently examined the role of electron-ion interactions on the stability of nuclei against electron captures and neutron emissions, 
both in the context of white dwarfs~\cite{cf15} and neutron stars~\cite{cfzh15,chamel2015b}, allowing for binary ionic mixtures and the 
presence of a strong magnetic field. In this paper, we pursue our investigation by taking into account the previously neglected effects 
of electron exchange and polarization. After discussing the stability condition in Section~\ref{stability}, applications to white dwarfs and neutron stars are presented in Section~\ref{applications}.

\section{Onset of electron capture and neutron emission by nuclei in dense plasmas}
\label{stability}

\begin{figure*}
\begin{center}
\includegraphics[scale=0.35]{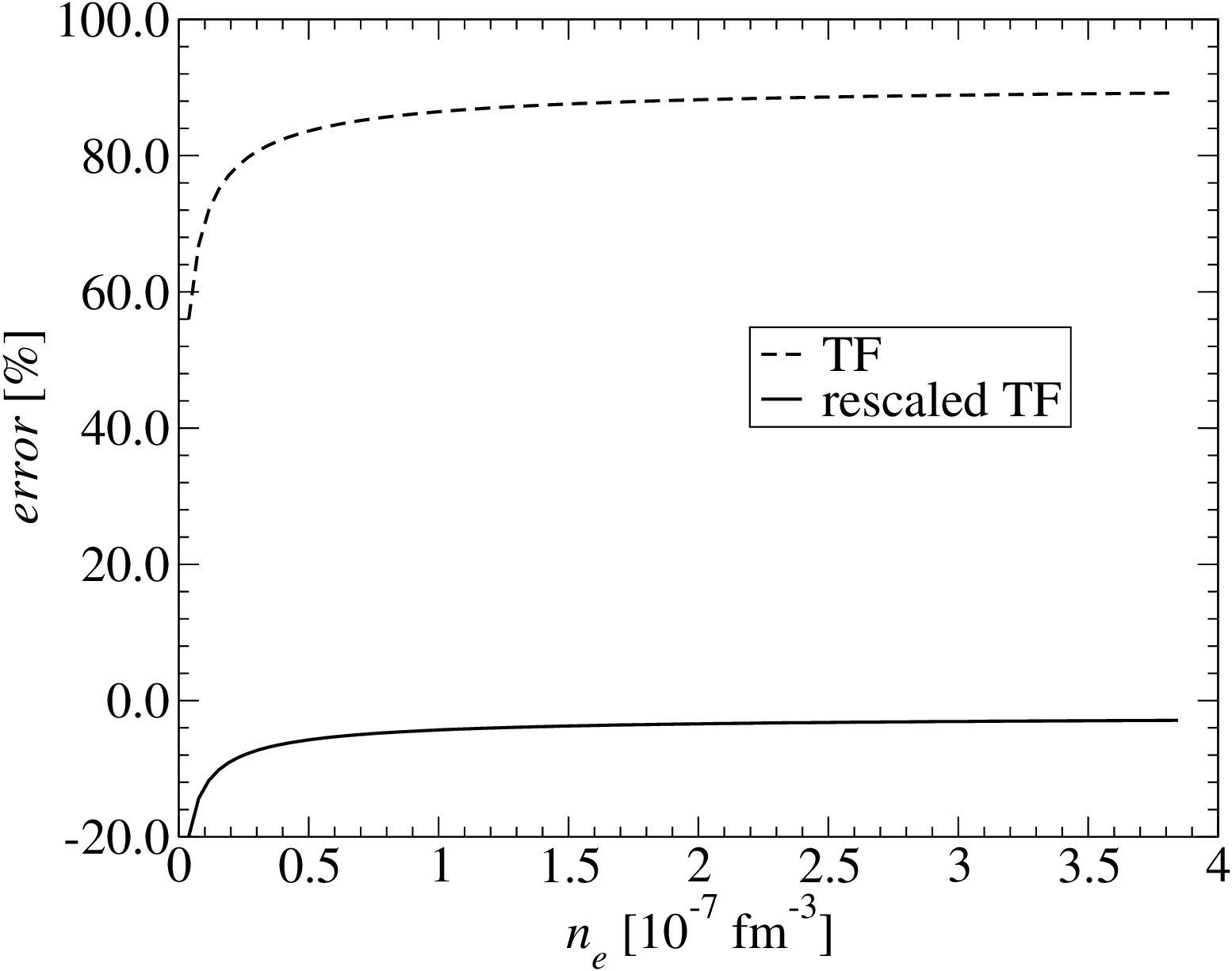}
\end{center}
\vskip -0.5cm
\caption{Relative deviation (in $\%$) between the electron polarization energy density obtained from Eq.~(42) of Potekhin and Chabrier~\cite{pot00} and 
two different approximations: the Thomas-Fermi (TF) approximation~(\ref{eq:epol-TF}) and the rescaled Thomas-Fermi expression~(\ref{eq:epol-approx}). 
The deviation is shown as a function of the electron number density for a cold dense Coulomb plasma of $^{56}$Fe. 
}
\label{fig1}
\end{figure*}

\begin{figure*}
\begin{center}
\includegraphics[scale=0.35]{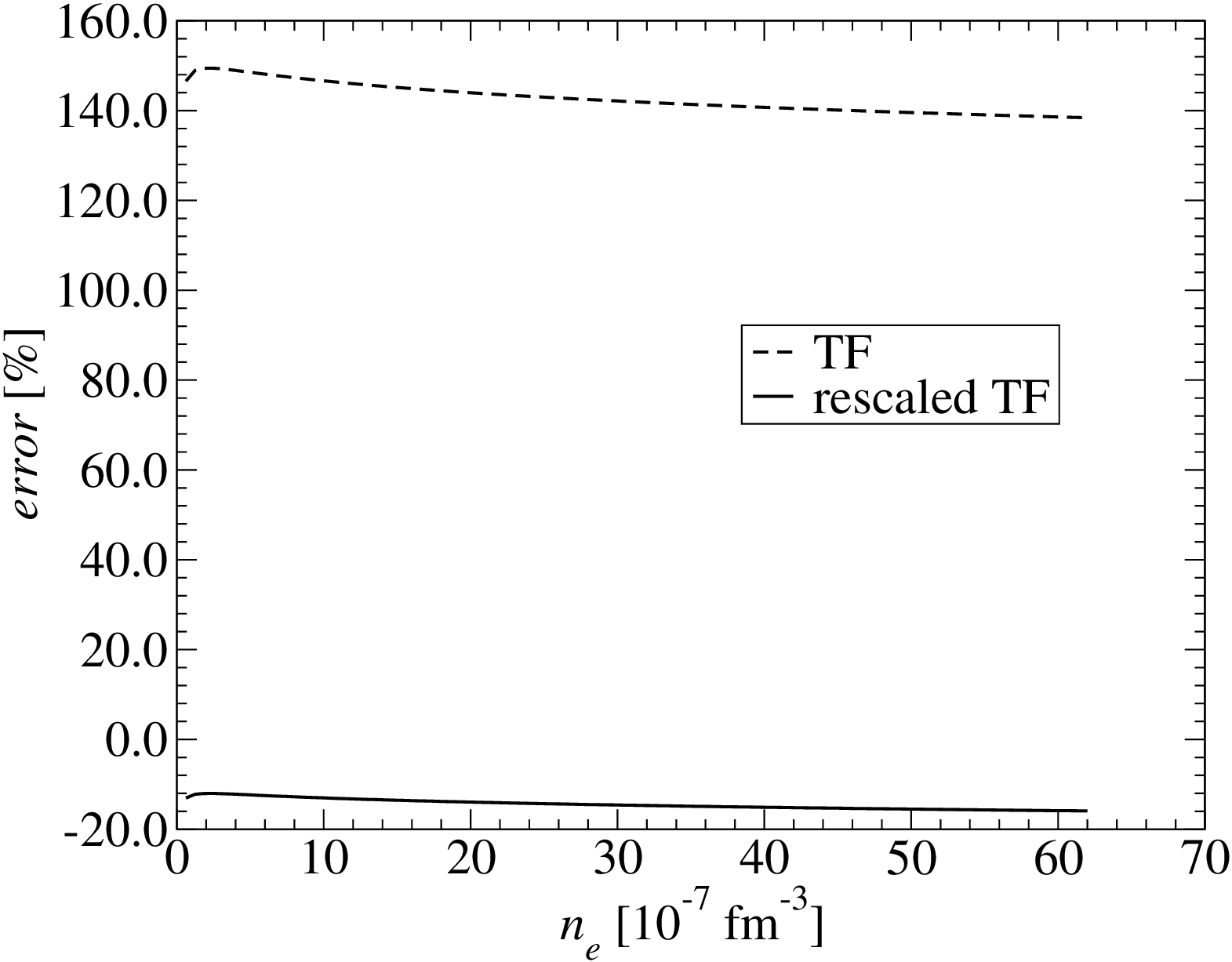}
\end{center}
\vskip -0.5cm
\caption{Same as Fig.~\ref{fig1} for a Coulomb plasma of $^{16}$O. 
}
\label{fig2}
\end{figure*}

\begin{figure*}
\begin{center}
\includegraphics[scale=0.35]{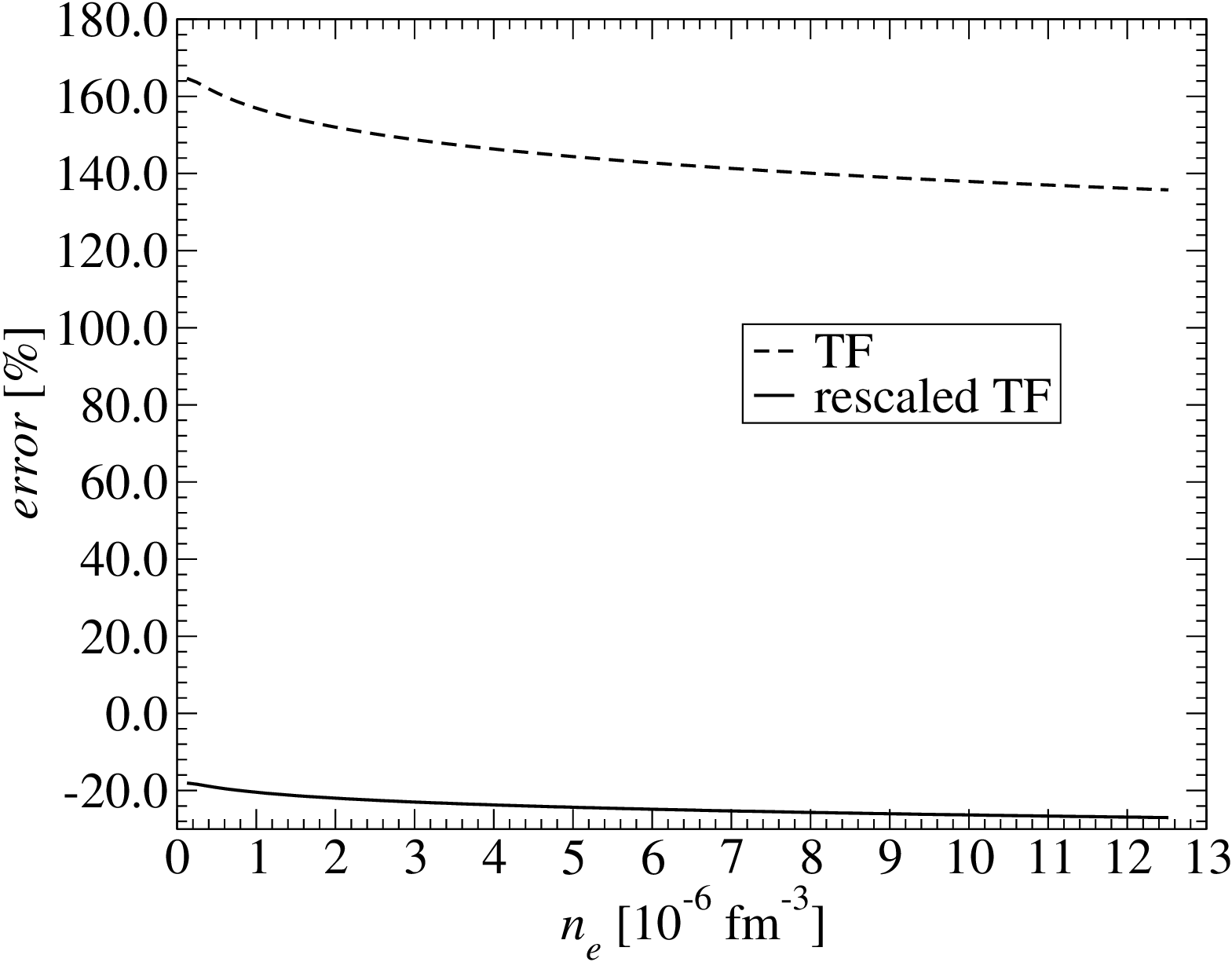}
\end{center}
\vskip -0.5cm
\caption{Same as Fig.~\ref{fig1} for a Coulomb plasma of $^{12}$C. 
}
\label{fig3}
\end{figure*}

As in our previous works~\cite{cf15,cfzh15,chamel2015b}, we consider matter at densities high enough that atoms are fully ionized. We further assume that
the temperature $T$ is lower than the crystallization temperature $T_m$ and that atomic nuclei are
arranged in a regular crystal lattice. For simplicity, we consider crystalline structures made of only
one type of ions  $^A_ZX$ with proton number $Z$ and mass number $A$. 
To a very good approximation, electrons can be treated as an ideal Fermi gas. 
The main correction arises from the electron-ion interactions. 
This model can be further refined by taking into account electron exchange and polarization effects. 
For ultrarelativistic electrons, the exchange contributions to the electron energy density and pressure are simply given 
by~\cite{salpeter61} 
\begin{equation}\label{eq:exc}
\frac{\mathcal{E}_e^\textrm{ex}}{\mathcal{E}_e}=\frac{P_e^\textrm{ex}}{P_e}=\frac{\alpha}{2\pi}\, ,
\end{equation}
where $\alpha=e^2/(\hbar c)$ is the fine structure constant ($e$ being the proton electric charge, $\hbar$ Planck-Dirac constant and $c$ the speed of light), $\mathcal{E}_e$ and $P_e$ are the energy density and the 
pressure of an ideal relativistic Fermi gas (see, e.g., Chap. 2 of Ref.~\cite{hae07}). 
For the electron polarization correction to the energy density, we shall use the interpolating formula (42) of Ref.~\cite{pot00}. 
We shall consider temperatures $T$ much lower than the plasma temperature $T_p=\hbar \omega_p/k_\textrm{B}$ ($k_\textrm{B}$ denoting Boltzmann's constant), where the ion-plasma 
frequency is given by $\omega_p=\sqrt{4\pi Z^2 e^2 n_i/M^\prime}$, $n_i$ is the ion number density, and 
$M^\prime$ denotes the ion mass (which coincides with the nuclear mass since atoms are fully ionized). The nuclear mass $M^\prime(A,Z)$ 
can be obtained from the corresponding tabulated atomic mass $M(A,Z)$ after substracting out the binding energy of the atomic electrons 
(see Eq.~(A4) of Ref.~\cite{lpt03}). 
Taking the limit $Z\rightarrow +\infty$, the electron polarization correction 
reduces to the familiar Thomas-Fermi expression~\cite{salpeter61}
\begin{equation}
\label{eq:epol-TF}
\mathcal{E}_e^\textrm{TF}=\frac{36}{35}\left(\frac{4}{9\pi}\right)^{1/3}\alpha Z^{2/3} \mathcal{E}_L\, ,
\end{equation}
written in terms of the lattice energy density $\mathcal{E}_L$, given by (see e.g. Chap. 2 of Ref.~\cite{hae07})
\begin{equation}\label{eq:EL}
\mathcal{E}_L=C e^2 n_e^{4/3} Z^{2/3}\, ,
\end{equation}
where $n_e$ is the electron number density, and the crystal structure constant $C$ is very-well approximated by the Wigner-Seitz estimate~\cite{salpeter61}
\begin{equation}\label{eq:WS-approx}
C=-\frac{9}{10}\left(\frac{4\pi}{3}\right)^{1/3}\, .
\end{equation}
For finite values of $Z$, and 
assuming $\Gamma_p \gg 1$, where 
\begin{equation}
\Gamma_p =\frac{Z^2 e^2}{a_i k_\textrm{B} T_p} \, ,
\end{equation}
and $a_i=Z^{1/3}(3/(4\pi n_e))^{1/3}$ is the ion sphere radius, the electron polarization correction for a body-centered cubic lattice 
can be approximately expressed as 
\begin{equation}\label{eq:epol-approx}
\mathcal{E}_{e}^\textrm{pol} \approx b_1(Z) \mathcal{E}_e^\textrm{TF}\, ,
\end{equation}
where the function $b_1(Z)$ is given by~\cite{pot00}
\begin{equation}
b_1(Z) = 1 - 1.1866 \,Z^{-0.267} + 0.27\,Z^{-1}\, .
\end{equation}
Using Eqs.~(\ref{eq:epol-TF}), (\ref{eq:EL}) and (\ref{eq:epol-approx}), the electron polarization contribution to the pressure is readily obtained 
\begin{equation}\label{eq:Ppol-approx}
P_{e}^\textrm{pol} \approx b_1(Z) \frac{\mathcal{E}_e^\textrm{TF}}{3}\, .
\end{equation}
Since $0< b_1\leq 1$, Eq.~(\ref{eq:epol-approx}) shows that the Thomas-Fermi approximation overestimates the electron 
polarization correction. In the crust of a neutron star, with proton numbers in the range $Z\sim 30-50$, we obtain 
$b_1\sim 0.5-0.6$. Because white dwarfs contain lighter elements, the deviations are significantly larger: $b_1 \sim 0.3-0.4$ 
for carbon and oxygen. As illustrated in Fig.~\ref{fig1}, Eq.~(\ref{eq:epol-approx}) provides a rather accurate approximation of the 
full expression from Potekhin and Chabrier~\cite{pot00} for the heavy elements expected to be found in neutron-star crusts. For the light elements contained 
in white dwarfs, the errors are substantially larger, as can be seen in Figs.~\ref{fig2} and \ref{fig3}. In all cases, we have plotted the 
electron polarization energy at densities below the onset of electron captures by nuclei, as explained below. In the following, we shall 
use the rescaled expressions~(\ref{eq:epol-approx}) and (\ref{eq:Ppol-approx}) for the electron polarization contribution to 
the energy density and pressure respectively. 

In the ultrarelativistic regime, 
the electron Fermi energy $\mu_e$ and the total pressure $P=P_e+P_L+P_e^\textrm{ex}+P_e^\textrm{pol}$ ($P_L=\mathcal{E}_L/3$ denotes the lattice contribution) can be approximately expressed as
\begin{equation}\label{eq:mue-ultra-rel}
\mu_e\approx \hbar c (3\pi^2 n_e)^{1/3}\gg m_e c^2\, ,
\end{equation}
\begin{equation}\label{eq:P-ultra-rel}
P\approx \frac{\mu_e^4}{12 \pi^2 (\hbar c)^3}\left(1+ \frac{\alpha}{2\pi} + \frac{4 C \alpha Z^{2/3}\sigma(Z)}{(81\pi^2)^{1/3}} \right) \, ,
\end{equation}
where $m_e$ is the electron mass. 
The term $\alpha/(2\pi)$ arises from electron exchange, while electron polarization effects are included in the dimensionless function
\begin{equation}\label{eq:sigma}
\sigma(Z)\equiv 1+\alpha \frac{ 12^{4/3}}{35 \pi^{1/3}}b_1(Z)  Z^{2/3}\, .
\end{equation}

With increasing compression, some fraction of nuclei $^A_ZX$ become unstable against electron captures accompanied by 
neutron emissions, and thus transform into nuclei $^{A-\Delta N}_{Z-\Delta Z}Y$ with the emission of $\Delta N$ free neutrons $n$ and $\Delta Z$ 
electron neutrinos $\nu_e$: 
\begin{equation}\label{eq:e-capture+n-emission}
^A_ZX+\Delta Z e^- \rightarrow ^{A-\Delta N}_{Z-\Delta Z}Y+\Delta N n+\Delta Z \nu_e\, .
\end{equation}
Following closely the analysis of Ref.~\cite{cfzh15} but accounting for electron exchange and polarization, we find that 
the stability condition is now given by  
\begin{eqnarray}\label{eq:e-capture+n-emission-gibbs-approx}
\mu_e\left(1+\frac{\alpha}{2\pi}\right)\Delta Z + C e^2 n_e^{1/3}\biggl[Z^{5/3}\sigma(Z)-(Z-\Delta Z)^{5/3}\sigma(Z-\Delta Z)
\nonumber \\
+ \frac{1}{3} Z^{2/3}\sigma(Z)\Delta Z\biggr] <  \mu_e^{\beta n} \, ,
\end{eqnarray}
\begin{equation}\label{eq:muebetan}
\mu_e^{\beta n}(A,Z)\equiv M^\prime(A-\Delta N,Z-\Delta Z)c^2-M^\prime(A,Z)c^2 +m_n c^2 \Delta N +  m_e c^2 \Delta Z  \, ,
\end{equation}
where $m_n$ denotes the neutron mass. 

In order to assess the range of validity of our zero-temperature treatment, let us estimate some characteristic temperatures. 
Approximating the electron Fermi energy by $\mu_e \approx \mu_e^{\beta n}/\Delta Z$, the electron Fermi temperature 
is simply given by 
\begin{equation}\label{TFe}
T_{\text{F}e}=\frac{\mu_e-m_e c^2}{k_\text{B}}\approx \frac{\mu_e^{\beta n}}{ k_\text{B}\Delta Z}\approx 5.93\times 10^9  \frac{\mu_e^{\beta n}}{m_e c^2 \Delta Z}~\text{K}\, .
\end{equation}
Using Eq.~(\ref{eq:mue-ultra-rel}), the crystallization temperature (see, e.g., Ref.~\cite{hae07}) can be expressed as 
\begin{equation}\label{Tm}
T_m=\frac{e^2 Z^2}{a_i k_\text{B} \Gamma_m}\approx \frac{\alpha \mu_e^{\beta n}}{k_\text{B} \Gamma_m\Delta Z}\left(\frac{4}{9\pi}\right)^{1/3} Z^{5/3}
\approx 1.29 \times 10^5 \frac{\mu_e^{\beta n}}{m_e c^2\Delta Z} Z^{5/3}~\text{K}\, ,
\end{equation}
where we have adopted the value $\Gamma_m\approx 175$ for the Coulomb coupling parameter at melting~\cite{hae07}. 
Likewise, the plasma temperature is given by 
\begin{equation}\label{Tp}
T_p=\frac{m_e c^2}{ k_\text{B}} \sqrt{\frac{4}{3\pi} \frac{m_e}{M^\prime}} \sqrt{\alpha Z}  \left(\frac{\mu_e^{\beta n}}{m_e c^2 \Delta Z}\right)^{3/2}
\approx 7.73\times 10^6 \sqrt{\frac{Z}{A}}  \left(\frac{\mu_e^{\beta n}}{m_e c^2 \Delta Z}\right)^{3/2}~\text{K}\, .
\end{equation}
We will provide numerical estimates of these temperatures for white dwarfs and neutron stars in the following Section. 

\section{Matter neutronization in compact stars}
\label{applications}

\subsection{White dwarfs} 

White dwarfs owe their existence to the presence of a highly degenerate electron gas in their interior, which provides
the necessary pressure to resist the gravitational collapse. The global stability of such stars can thus be limited by the onset 
of electron captures (\ref{eq:e-capture+n-emission}) with $\Delta Z=1$ and $\Delta N=0$ (see, e.g. Ref.~\cite{cf15}). 
Solving Eq.~(\ref{eq:e-capture+n-emission-gibbs-approx}) 
in the limit of ultrarelativistic electrons, the average baryon density and pressure for the onset of electron capture are 
approximately given by
\begin{eqnarray}\label{eq:rhobeta}
 n_\beta(A,Z) &&\approx \frac{A}{Z} \frac{\mu_e^\beta(A,Z)^3}{3\pi^2 (\hbar c)^3}\biggl[1+\frac{\alpha}{2\pi}\nonumber \\
 &&+\frac{C \alpha}{(3\pi^2)^{1/3}}\left(Z^{5/3}\sigma(Z)-(Z-1)^{5/3}\sigma(Z-1)+\frac{Z^{2/3}\sigma(Z)}{3}\right)\biggr]^{-3}\, ,
\end{eqnarray}
\begin{eqnarray}\label{eq:Pbeta}
P_\beta(A,Z) &\approx& \frac{\mu_e^\beta(A,Z)^4}{12 \pi^2 (\hbar c)^3}\biggl[1+\frac{\alpha}{2\pi}+\frac{4C \alpha Z^{2/3}\sigma(Z)}{(81\pi^2)^{1/3}} \biggr]\\
 \nonumber
&&\times \biggl[1+\frac{\alpha}{2\pi}+\frac{C \alpha}{(3\pi^2)^{1/3}}\left(Z^{5/3}\sigma(Z)-(Z-1)^{5/3}\sigma(Z-1)+\frac{Z^{2/3}\sigma(Z)}{3}\right)\biggr]^{-4}\, .
\end{eqnarray}
\begin{equation}\label{eq:muebeta}
\mu_e^\beta(A,Z)= M^\prime(A,Z-1)c^2-M^\prime(A,Z)c^2+ m_e c^2 \, .
\end{equation}
Equations~(\ref{eq:rhobeta}) and (\ref{eq:Pbeta}) generally represent the highest density and pressure that can be found in white dwarfs 
(the actual values of the central density and pressure may be lower due to general relativity, see e.g. Ref.~\cite{shapiro1983}). As an example, we
consider a stellar core made of $^{16}$O. Using the masses from the 2012 Atomic Mass Evaluation~\cite{audi2012}, we find $\mu_e^\beta=10.931$~MeV. 
Substituting in Eqs.~(\ref{eq:rhobeta}) and (\ref{eq:Pbeta}) considering a body-centered cubic lattice with $C = -1.444231$~\cite{bcc}, we obtain 
$n_\beta=1.240\times 10^{-5}$ fm$^{-3}$
($n_\beta=1.244\times 10^{-5}$ fm$^{-3}$) and $P_\beta=1.709\times 10^{-5}$~MeV~fm$^{-3}$ ($P_\beta=1.714\times 10^{-5}$~MeV~fm$^{-3}$) with (without) 
electron exchange and polarization corrections. The corresponding characteristic temperatures given by Eqs.~(\ref{TFe}), (\ref{Tm}), and 
(\ref{Tp}) are respectively $T_{\text{F}e}=1.6\times 10^{10}$~K, $T_m=1.1\times 10^7$~K, and $T_p=4.2\times 10^7$~K. The assumption of crystallized 
white-dwarf cores therefore yields the most stringent condition on the highest temperature up to which the present treatment is valid. Finally, 
the plasma parameter $\Gamma_p$, which in the ultrarelativistic regime $\mu_e \gg m_e c^2$ can be approximately expressed as 
\begin{equation}\label{eq:Gammap_app}
\Gamma_p \approx \sqrt{\frac{\alpha M^\prime c^2}{\mu_e}} \left(\frac{\pi}{12}\right)^{1/6} Z^{7/6}
\end{equation}
with $\mu_e\approx \mu_e^\beta/\Delta Z$, 
is given by $\Gamma_p=29$ for $^{16}$O. The condition $\Gamma_p \gg 1$ underlying Eqs.~(\ref{eq:epol-approx}) and (\ref{eq:Ppol-approx}) is thus 
well satisfied.  

\subsection{Nonaccreting neutron stars}

As discussed in Ref.~\cite{cfzh15}, the onset of neutron drip can be determined from Eqs.~(\ref{eq:e-capture+n-emission-gibbs-approx}) and (\ref{eq:muebetan}) 
with $\Delta Z=Z$ and $\Delta N=A$.  Solving Eq.~(\ref{eq:e-capture+n-emission-gibbs-approx}) in the limit of ultra relativistic electrons, the average baryon 
density and pressure for the onset of neutron drip are approximately given by
\begin{equation}\label{eq:rhodrip-multi}
n_{\rm drip}(A,Z) \approx \frac{A}{Z} \frac{\mu_e^{\rm drip}(A,Z)^3}{3\pi^2 (\hbar c)^3}
\biggl[1+\frac{\alpha}{2\pi} + \frac{4 C \alpha}{(81\pi^2)^{1/3}} Z^{2/3}\sigma(Z) \biggr]^{-3}\, ,
\end{equation}
\begin{equation}\label{eq:Pdrip-multi}
P_\textrm{drip}(A,Z) \approx \frac{\mu_e^\textrm{drip}(A,Z)^4}{12 \pi^2 (\hbar c)^3}
\biggl[1+\frac{\alpha}{2\pi}+\frac{4C \alpha Z^{2/3}\sigma(Z)}{(81\pi^2)^{1/3}} \biggr]^{-3} \, ,
\end{equation}
\begin{equation}\label{eq:muedrip}
\mu_e^{\textrm{drip}}(A,Z)\equiv \frac{-M^\prime(A,Z)c^2+A m_n c^2}{Z} +m_e c^2 \, .
\end{equation}
In this case, the effect of electron polarization is to replace the proton number $Z$ in the lattice term by   
an \emph{effective} atomic number, defined by 
\begin{equation}\label{eq:zeff}
Z_{\rm eff}\equiv Z \sigma(Z)^{3/2}=Z \left(1+\alpha \frac{12^{4/3}}{35 \pi^{1/3}} b_1(Z) Z^{2/3}\right)^{3/2}> Z\, .
\end{equation}

\begin{figure*}
\begin{center}
\includegraphics[scale=0.45]{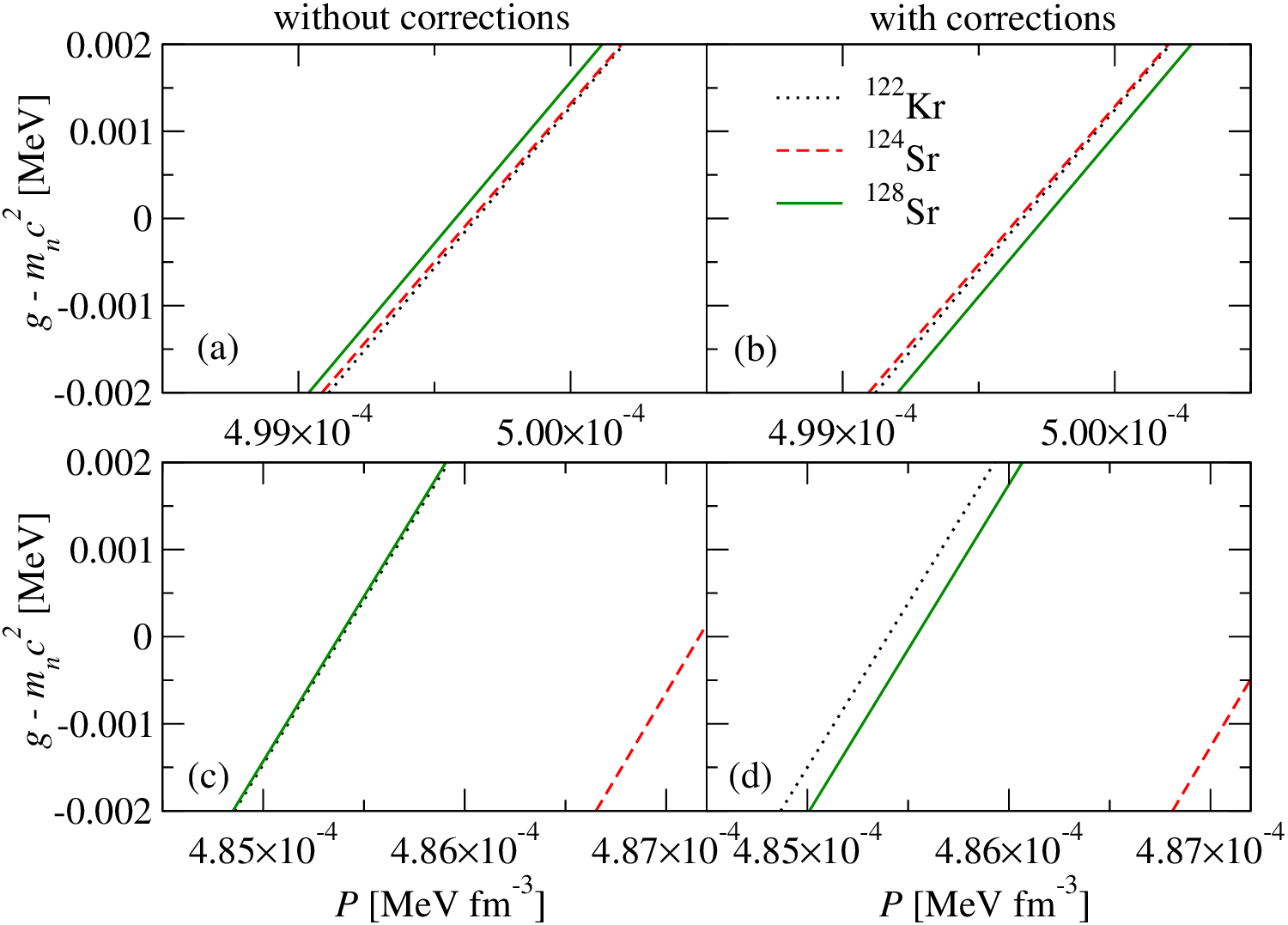}
\end{center}
\caption{Gibbs free energy per nucleon (with the neutron mass energy subtracted out) versus pressure for the HFB-22 (panels a and b) and HFB-24 (panels c and d) mass 
models, for nonaccreting neutron stars, for three different nuclei. See text for details.}
\label{fig:hfb22-24-g}
\end{figure*}

Although electron exchange and polarization corrections are small, they may change the composition. 
The equilibrium nucleus at pressure $P$ is determined by minimizing the Gibbs free energy per nucleon defined by 
\begin{equation}
\label{eq:gibbs-general}
g = \frac{\mathcal{E}+P}{n} \ ,
\end{equation}
where $\mathcal{E}$ denotes the mean energy density of matter, and $n$ is the mean baryon number density. 
Ignoring electron exchange and polarization effects, Eq.~(\ref{eq:gibbs-general}) can be equivalently expressed 
as~\cite{bps1971} 
\begin{equation}
\label{eq:gibbs0}
g=\frac{M^\prime(A,Z)c^2}{A}+\frac{Z}{A}\left(\mu_e -m_e c^2+\frac{4}{3} C e^2 n_e^{1/3} Z^{2/3}\right)\, .
\end{equation}
Using the general definition~(\ref{eq:gibbs-general}) as well as Eqs.~(\ref{eq:exc}), (\ref{eq:epol-TF}),  
(\ref{eq:epol-approx}), and (\ref{eq:Ppol-approx}), it can be easily seen that including electron exchange and 
polarization corrections yields 
\begin{equation}
\label{eq:gibbs}
g=\frac{M^\prime(A,Z)c^2}{A}+\frac{Z}{A}\biggl[\mu_e\left(1+\frac{\alpha}{2\pi}\right) -m_e c^2+\frac{4}{3} C e^2 n_e^{1/3} Z_\textrm{eff}^{2/3}\biggr]\, .
\end{equation}
The nuclear masses of relevance for the crust region of interest here have not yet been measured (see, e.g. Refs.~\cite{roca2008,pearson2011,wolf2013,kreim2013}). 
For this reason, we have made use of the latest microscopic mass tables taken from the BRUSLIB database~\cite{bruslib}. These masses were computed by the Brussels-Montreal 
group and are based on either generalized~\cite{goriely2013} or standard Skyrme forces~\cite{goriely2013b}. 
These models fit the 2353 measured masses of nuclei with $N$ and $Z \geq 8$ from the 2012 Atomic Mass Evaluation \cite{audi2012}, with a root-mean-square deviation 
of about $0.6$~MeV and $0.5$~MeV, respectively. For comparison, we have also employed the microscopic mass table based on the Gogny force D1M~\cite{d1m}, as well as 
the more phenomenological model of Duflo and Zucker~\cite{duflo1995}. 
We have assumed that nuclei are arranged on a body-centered cubic lattice. 
Results are summarized in Table~\ref{tab:drip-cat}. Characteristic temperatures for the validity of our crust model are given in Table~ \ref{tab:drip-cat-temp}. 
The validity of our treatment of electron polarization at finite temperatures is limited by the plasma temperature $T_p\sim 10^9$~K. This corresponds to an effective surface temperature (as seen by an observer at infinity) between 
$10^{6.6}$~K and $10^{6.8}$~K (see Fig.7 of Ref.~\cite{page06}). Depending on the cooling scenario, our model is thus applicable to neutron stars whose 
estimated age ranges from a few months up to several million years~\cite{gne01}. 
The plasma parameter $\Gamma_p$, which is approximately given by Eq.~(\ref{eq:Gammap_app}) with $\mu_e\approx \mu_e^{\rm drip}$, is of the order 
of $\Gamma_p\sim 3\times 10^2$. Therefore the electron polarization formulas of Potekhin and Chabrier~\cite{pot00} can be very well approximated by Eqs.~(\ref{eq:epol-approx}) and (\ref{eq:Ppol-approx}).  
As for the analytical expressions~(\ref{eq:rhodrip-multi}) and (\ref{eq:Pdrip-multi}), the errors amount to about $0.2\%$ at most. 
The role of electron exchange and polarization effects to the neutron-drip transition is  found to be negligible ($\sim 0.1\%$) for all models but 
HFB-22~\cite{goriely2013}. Although the neutron-drip density and pressure predicted by this particular model are only slightly shifted (by less than $0.3\%$), 
the equilibrium nucleus changes from $^{122}$Kr to $^{128}$Sr. This latter result stems from the fact that the HFB-22 nuclear mass model predicts very similar 
values for the threshold electron Fermi energy $\mu_e^{\rm drip}$ for nuclei $^{128}$Sr and $^{122}$Kr: $24.967$ and $25.004$~MeV respectively. As a consequence, 
even small corrections to the Gibbs free energy per nucleon $g$ can change the minimum. In order to better illustrate this point, we have shown in 
Fig.~\ref{fig:hfb22-24-g} the Gibbs free energy per nucleon (with the neutron mass energy subtracted out) around the neutron-drip pressure, as obtained using the HFB-22 
(upper panels) and HFB-24 (lower panels) mass models. The values of $g - m_n c^2$ 
are shown for three different nuclei (see Table~\ref{tab:drip-cat}). The equilibrium nucleus at the neutron-drip transition is the one minimizing $g$, and such 
that $g - m_n c^2 = 0$. 
In the right (left) panels, we show the results obtained with (without) including electron exchange and polarization corrections. 
For the model HFB-22 (upper panels), the differences in the Gibbs free energies for the three nuclei are so small that including or not electron exchange and 
polarization corrections leads to different equilibrium nuclei. For the model HFB-24 (lower panels), the Gibbs free energy per nucleon for  $^{124}$Sr is much 
lower than for any other nuclei so that electron exchange and polarization corrections do not change the composition. 

At the temperatures $T < T_p\sim 10^9$~K considered here, thermal effects are not expected to play a role, since thermal corrections are exponentially suppressed by shell and 
pairing effects. Indeed, as shown in Ref.~\cite{onsi08}, the composition of neutron-star crusts at the neutron-drip transition remains unchanged up to temperatures 
of the order of $10^9$~K.

The peculiar prediction of model HFB-22 
could be purely accidental. As a matter of fact, in the recent series of Brussels-Montreal mass models \cite{goriely2013}, HFB-22 (HFB-24) was found to be in 
worst (best) agreement with various constraints coming from both nuclear physics and astrophysics~\cite{pearson2014, fantina2015}. More importantly, corrections 
due to electron exchange and polarization are found to be much smaller than the uncertainties 
related to nuclear masses, as can be seen in Table~\ref{tab:drip-cat}.

\begin{table}
\centering
\caption{Neutron-drip transition in the crust of nonaccreting neutron stars, as predicted by different nuclear mass models: 
mass and atomic numbers of the dripping nucleus, baryon number density and corresponding pressure, electron Fermi energy. 
Values in parentheses are calculated without including electron exchange and polarization. See text for details.}\smallskip
\label{tab:drip-cat}
\begin{tabular}{cccccc}
\hline 
 & $A$ & $Z$ & $n_\textrm{drip}$ ($10^{-4}$ fm$^{-3}$) & $P_\textrm{drip}$ ($10^{-4}$ MeV~fm$^{-3}$) & $\mu_e^{\rm drip}$ (MeV)\\
\hline \noalign {\smallskip}
HFB-22 & 128 (122) & 38 (36) & 2.701 (2.707) & 5.004 (4.989) & 24.97 (25.00)\\ 
HFB-24 & 124 & 38 & 2.565 (2.564) & 4.871 (4.871) & 24.81\\    
HFB-27 & 124 & 38 & 2.547 (2.547) & 4.828 (4.828) & 24.76 \\ 
D1M & 120 & 38 & 2.454 (2.454) & 4.799 (4.799) & 24.72  \\
DZ & 118 & 36 & 2.578 (2.578) & 4.886 (4.886) & 24.87  \\
\hline
\end{tabular}
\end{table}

\begin{table}
\centering
\caption{Characteristic temperatures at the neutron-drip transition in the crust of nonaccreting neutron stars, as predicted by different 
nuclear mass models: electron Fermi temperature $T_{{\rm F}e}$, crystallization temperature $T_m$, and plasma temperature $T_p$.}\smallskip
\label{tab:drip-cat-temp}
\begin{tabular}{cccc}
\hline 
 & $T_{{\rm F}e}$ (K) & $T_m$ (K) & $T_p$ (K) \\
\hline \noalign {\smallskip}
HFB-22 & $2.9\times 10^{11}$ ($2.9\times 10^{11}$) & $2.7\times 10^9$ ($2.5\times 10^9$) & $1.4\times 10^9$ ($1.4\times 10^9$)  \\
HFB-24 & $2.9\times 10^{11}$ & $2.7\times 10^9$ & $1.4\times 10^9$  \\
HFB-27 & $2.9\times 10^{11}$ & $2.7\times 10^9$ & $1.4\times 10^9$  \\
D1M & $2.9\times 10^{11}$ & $2.7\times 10^9$ & $1.5\times 10^9$  \\
DZ & $2.9\times 10^{11}$ & $2.5\times 10^9$ & $1.4\times 10^9$  \\
\hline
\end{tabular}
\end{table}

\subsection{Accreting neutron stars}

As discussed in Ref.~\cite{cfzh15}, the onset of neutron drip can be determined from Eqs.~(\ref{eq:e-capture+n-emission-gibbs-approx}) and (\ref{eq:muebetan}) with
$\Delta Z=1$ and $\Delta N>0$. Solving Eq.~(\ref{eq:e-capture+n-emission-gibbs-approx}) in the limit of ultra relativistic electrons, the average baryon 
density and pressure for the onset of neutron drip are approximately given by
\begin{eqnarray}\label{eq:rhodrip-acc}
 n_\textrm{drip-acc}(A,Z) &&\approx \frac{A}{Z} \frac{\mu_e^\textrm{drip-acc}(A,Z)^3}{3\pi^2 (\hbar c)^3}\biggl[1+\frac{\alpha}{2\pi}\nonumber \\
 &&+\frac{C \alpha}{(3\pi^2)^{1/3}}\left(Z^{5/3}\sigma(Z)-(Z-1)^{5/3}\sigma(Z-1)+\frac{Z^{2/3}\sigma(Z)}{3}\right)\biggr]^{-3}\, ,
\end{eqnarray}
\begin{eqnarray}\label{eq:Pdrip-acc}
P_\textrm{drip-acc}(A,Z) &\approx& \frac{\mu_e^\textrm{drip-acc}(A,Z)^4}{12 \pi^2 (\hbar c)^3}\biggl[1+\frac{\alpha}{2\pi}
+\frac{4C \alpha Z^{2/3}\sigma(Z)}{(81\pi^2)^{1/3}} \biggr]\\ \nonumber
&&\times \biggl[1+\frac{\alpha}{2\pi}+\frac{C \alpha}{(3\pi^2)^{1/3}}\left(Z^{5/3}\sigma(Z)-(Z-1)^{5/3}\sigma(Z-1)+\frac{Z^{2/3}\sigma(Z)}{3}\right)\biggr]^{-4}\, .
\end{eqnarray}
\begin{equation}\label{eq:muedrip-acc}
\mu_e^\textrm{drip-acc}(A,Z)= M^\prime(A-\Delta N,Z-1)c^2-M^\prime(A,Z)c^2+\Delta N m_n c^2 + m_e c^2 \, .
\end{equation}
The changes of the neutron-drip density and pressure due to electron exchange and polarization are found to be negligibly small for all models: deviations are of the 
order of $\sim 0.1\%$. 

As explained in Ref.~\cite{cfzh15}, the dripping nucleus $^A_ZX$ is such that the capture of an electron with the emission of free neutrons
($\Delta Z=1$, $\Delta N>0$) has a lower threshold energy than the capture alone ($\Delta Z=1$, $\Delta N=0$) , i.e. $\mu_e^{\beta n}(A,Z) < \mu_e^\beta(A,Z)$. 
The proton number $Z$ of the dripping nucleus is thus the highest proton number lower than that of the initial ashes and for which the $\Delta N$-neutron 
separation energy defined as
\begin{equation}
\label{eq:sn-dn}
S_{\Delta N n}(A,Z-1) \equiv M(A-\Delta N,Z-1)c^2 - M(A,Z-1)c^2 + \Delta N m_n c^2
\end{equation}
is negative. This condition depends only on nuclear masses. As a consequence, electron exchange and polarization do \emph{not} change the dripping nucleus in 
accreting neutron-star crusts. The composition found in Ref.~\cite{cfzh15} thus remains unchanged. 

\section{Conclusions}

The electron polarization expressions proposed by Potekhin and Chabrier are found to deviate substantially from the Thomas-Fermi approximation that has been widely 
employed in studies of dense Coulomb plasmas in compact stars. On the other hand, for most astrophysical purposes these deviations can be fairly accurately taken 
into account by merely rescaling the Thomas-Fermi expression for the correction to the density and pressure, Eqs.(\ref{eq:epol-approx}) and (\ref{eq:Ppol-approx}) 
respectively. In particular, the Thomas-Fermi results have to be reduced by about a factor $\sim 2-3$ for the elements likely to be present in white-dwarf cores 
and neutron-star crusts. 
Using this rescaled Thomas-Fermi expression, we have studied the importance of electron polarization effects on the onset of electron captures and neutron 
emissions by nuclei in white-dwarf cores and neutron-star crusts. We have also taken into account electron exchange since the corresponding corrections to the 
energy density and pressure are generally of the same order of magnitude as those due to  electron polarization. We have extended the instability condition for
 the onset of electron captures and neutron emissions by nuclei so as to include electron exchange and polarization. The new condition is embedded in 
Eq. (\ref{eq:e-capture+n-emission-gibbs-approx}). We have derived analytical expressions for the threshold density and pressure at the onset of electron captures 
in white-dwarf cores, namely Eqs.~(\ref{eq:rhobeta}) and (\ref{eq:Pbeta}). The corrections due to electron exchange and polarization effects are very small, 
about $0.3\%$ for cores made of $^{16}$O. 
Moreover, we have obtained analytical expressions for the neutron-drip density and pressure in accreting and nonaccreting neutron-star crusts, see 
Eqs.~(\ref{eq:rhodrip-multi}), (\ref{eq:Pdrip-multi}), (\ref{eq:rhodrip-acc}), and (\ref{eq:Pdrip-acc}). We have determined the composition of nonaccreting 
neutron-star crusts by minimizing numerically the Gibbs free energy per nucleon. 
For the nuclear masses that have not yet been measured, we have used the predictions from different models: the latest Skyrme-Hartree-Fock-Bogoliubov mass 
tables computed by the Brussels-Montreal group~\cite{goriely2013,goriely2013b} available on the BRUSLIB database~\cite{bruslib}, the Hartree-Fock-Bogoliubov mass table 
based on the Gogny force D1M \cite{d1m}, and the more phenomenological model of Duflo and Zucker \cite{duflo1995}. In this way, we have also calculated 
numerically the neutron-drip density and pressure. The precision of the analytical formulas corresponds to an error of about $0.2\%$ at most. The neutron-drip 
density and pressure are hardly changed by electron exchange and polarization effects. The deviations lie below $0.3\%$. For the model HFB-22, the equilibrium 
nucleus changes from $^{122}$Kr to $^{128}$Sr thus leading to a slightly larger shift of the neutron-drip density and pressure than for the other models.
 This change of composition stems from the fact that HFB-22 predicts very similar values 
for the threshold electron Fermi energy $\mu_e^\textrm{drip}$ for nuclei $^{128}$Sr and $^{122}$Kr: $24.967$ and $25.004$~MeV respectively. In turn, 
$\mu_e^\textrm{drip}$ depends solely on the mass of the neutron-drip nucleus. The current uncertainties in nuclear masses are found to be more important than
 electron exchange and polarization effects. The composition of the outer crust of neutron stars towards the neutron-drip point thus requires a more accurate 
determination of the masses of Kr and Sr neutron-rich isotopes.

\begin{acknowledgments}
This work has been mainly supported by Fonds de la Recherche Scientifique - FNRS (Belgium). 
Partial support comes also from the European Cooperation in Science and Technology (COST) Action \emph{NewCompStar} MP1304 . 
\end{acknowledgments}

\end{document}